\documentclass[superscriptaddress,twocolumn,showpacs,a4paper,
amssymb,amsmath,nobibnotes,aps,prd,
showkeys,
nofootinbib,notitlepage]{revtex4-1}
\usepackage{verbatim}
\usepackage[T1]{fontenc}
\usepackage[utf8]{inputenc}
\usepackage[american]{babel}
\usepackage{epsfig}
\usepackage{graphicx,subcaption,caption}
\usepackage{booktabs}
\usepackage{multirow}
\usepackage{dcolumn}
\usepackage{amsmath}
\usepackage{mathtools}
\usepackage{amsfonts}
\usepackage{amssymb}
\usepackage{epstopdf}
\usepackage{bm}
\usepackage{siunitx}
\usepackage{braket}
\usepackage{enumitem}
\usepackage{soul}
\usepackage[table]{xcolor}
\usepackage{color}
\usepackage{transparent}
\usepackage[
font=footnotesize,justification=justified]{caption}
\usepackage{pifont}

\definecolor{navyblue}{rgb}{0.0, 0.0, 0.5}
\definecolor{royalblue}{rgb}{0.25, 0.41, 0.88}
\definecolor{cadmiumgreen}{rgb}{0.0, 0.42, 0.24}
\definecolor{blue-violet}{rgb}{0.54, 0.17, 0.89}
\definecolor{darkviolet}{rgb}{0.58, 0.0, 0.83}
\definecolor{orange(colorwheel)}{rgb}{1.0, 0.5, 0.0}

\usepackage{hyperref}
\hypersetup{
	colorlinks=true, 
	linkcolor=royalblue, 
	citecolor=magenta}

\usepackage{booktabs}
\usepackage{multirow}
\usepackage{dcolumn}
\usepackage{colortbl}

\begin{document}
	
    \title{Higher-Curvature Corrections and Tensor Modes}
    
	\author{William Giarè}
	\email{william.giare@uniroma1.it}
	\affiliation{Physics Department and INFN, Universit\`a di Roma ``La Sapienza'', Ple Aldo Moro 2, 00185, Rome, Italy}

	\author{Fabrizio Renzi}
	\email{renzi@lorentz.leidenuniv.nl}
	\affiliation{%
		Lorentz Institute for Theoretical Physics, Leiden University, PO Box 9506, Leiden 2300 RA, The Netherlands
	}%
	
	\author{Alessandro Melchiorri}
	\email{alessandro.melchiorri@roma1.infn.it}
	\affiliation{Physics Department and INFN, Universit\`a di Roma ``La Sapienza'', Ple Aldo Moro 2, 00185, Rome, Italy}

\date{\today}
\preprint{}
\begin{abstract}
Higher-curvature corrections to the effective gravitational action may leave signatures in the spectrum of primordial tensor perturbations if the inflationary energy scale is sufficiently high. In this paper we further investigate the effects of a coupling of the Inflaton field to higher-curvature tensors in models with a minimal breaking of conformal symmetry. We show that an observable violation of the tensor consistency relation from higher-curvature tensors implies also a relatively large running of the tensor tilt, enhanced even by some order of magnitude with respect to the standard slow roll case. This may leave signatures in the tensor two-point function that we could test to recognize higher-curvature effects, above all if they are translated into a blue tilted spectrum visible by future Gravitational Wave experiments. Exploiting current cosmic microwave background and gravitational wave data we also derive constraints on the inflationary parameters, inferring that large higher-curvature corrections seem to be disfavored.

\end{abstract}


\maketitle

\section{Introduction}

In the very early Universe a phase of accelerated expansion known as Cosmological Inflation is expected to solve the flatness and horizon problem, setting the initial condition for Hot Big Bang Theory evolution~\cite{Guth:1980zm}. 

The inflationary Universe is close to de Sitter, a maximally symmetric solution of the Einstein equations with a positive cosmological constant. It is well known that the de Sitter spacetime, being maximally symmetric, has 10 different Killing vectors ( \textit{i.e.} the maximum possible number for a 4-dimensional spacetime) that roughly correspond to 10 different isometries, namely: 3 spatial translations, 3 spatial rotations, 1 dilatation and 3 special conformal transformations\footnote{At late times, special conformal transformations act like conformal transformations on the space-like boundary~\cite{Hawking:1973uf,Baumann:2015xxa}.} \cite{Hawking:1973uf,Spradlin:2001pw,Baumann:2015xxa,Kehagias:2015jha,Kehagias:2013xga,Kehagias:2017rpe,Anninos:2019nib,Lee:2017okg}. However in almost any physical model of inflation the de Sitter symmetries are broken to ensure the end of inflation and the simplest way is assuming a dynamical scalar field $\phi$, the Inflaton, driving inflation. Indeed a dynamical scalar field introduces a time dependence vacuum expectation value or equivalently a preferred time slicing of the de Sitter spacetime, basically providing a well defined "clock" for measuring the time to the end of inflation~\cite{Weinberg:2008hq,Lyth:1998xn,Baumann:2009ds,Baumann:2015xxa,Baumann:2014nda,Cheung:2007st,Lee:2017okg}.

Moreover the inflationary vacuum fluctuations, becoming classical on large scales, can induce energy-density fluctuations, sourcing both rotational invariant scalar perturbations and tensor perturbations with helicity $\pm2$, the so-called Primordial Gravitational Waves (PGWs). Scalar and Tensor perturbations are decoupled at the linearized level: after the end of Inflation, during the radiation dominated era, scalar perturbations reenter the observable Universe, providing the seeds for the structure formation and giving a natural explanation for the observed anisotropies of the Cosmic Microwave Background (CMB). PGWs may instead produce fluctuations in the polarization of the CMB photons, leading to a very distinctive signature in the CMB B-modes polarization power spectrum on large angular scale~\cite{Guth:1980zm,Starobinsky:1980te,Linde:1981mu,Vilenkin:1983xq,Mukhanov:2005sc,Dodelson:2003ft,Weinberg:2008zzc,Martin:2013tda,Baumann:2009ds,Clarke:2020bil}.  The power spectra of scalar and tensor perturbations in a quasi de Sitter geometry are expected to be nearly but not exactly flat since they acquire a small scale dependence quantified by the same slow roll parameter $\epsilon_{1}\doteq - \dot H / H^2 \simeq \dot \phi^2 / (2M_p^2\,H^2)$ that controls also the breaking of the conformal symmetry, restored in the limit $\epsilon_1\to 0$. In the simplest single field slow roll inflation minimally coupled to gravity, the spectrum of tensor fluctuations is characterized by the well known consistency relation $r=16\epsilon_1=-8\,n_{\rm T}$ between the amplitude (parametrized through the so called tensor to scalar ratio $r$) and the tilt $n_{\rm T}$, with the inflationary energy scale itself proportional to $r$ \cite{Sahni:1990tx, Lyth:2009zz,Mukhanov:2005sc,Dodelson:2003ft,Weinberg:2008zzc,Martin:2013tda,Kamionkowski:2015yta,Mirbabayi:2014jqa,Ozsoy:2014sba,Lee:2017okg,Baumann:2009ds,Kinney:2009vz,Renzi:2019ewp}. While both the amplitude and the tilt of the scalar spectrum are measured with good precision~\cite{Akrami:2018odb}, a detection of primordial tensor modes is still missing and a combined analysis of the current Planck \cite{Akrami:2018odb} and BICEP2/Keck array (BK15)~\cite{Ade:2018gkx} data only sets an upper bound $r_{0.002} < 0.056$ at $95\%$ C.L on the tensor amplitude. Nevertheless, in the upcoming decade a new generation of CMB experiments (\textit{e.g.} BICEP3~\cite{BICEP3}, CLASS~\cite{CLASS} , SPT-3G~\cite{SPT-3G}, Advanced ACTPol~\cite{ACTPol}, LBIRD~\cite{LBIRD} and CMB-S4~\cite{CMB-S4}) is expected to reach a sensitivity $r \sim 0.01 - 0.001$ possibly leading to a first detection of tensor modes for sufficiently high-scale inflation. This may open up the possibility of probing physics at extremely high energy scale, for example testing deviations from the standard inflationary predictions as a hint for new physics.

If the inflationary energy scale is sufficiently high, higher-curvature corrections to the gravitational effective action, expected for example in string theory~\cite{Zwiebach:1985uq,Baumann:2014nda,Gasperini:1997up,Cai:2015ipa,Gasperini:2007vw}, can lead to testable features in the primordial perturbations~\cite{Baumann:2015xxa,Pajer:2016ieg,Creminelli:2011mw,Satoh:2008ck,Gasperini:1997up,Gasperini:2007vw,Cai:2015ipa,Mishima:2019vlh,Yi:2018gse,Wu:2017joj,Koh:2016abf,Hikmawan:2015rze,Jiang:2013gza,Watanabe:2010fh,Guo:2009uk,Edelstein:2020lgv,Alvarez-Gaume:2015rwa,Dalianis:2014aya, Oikonomou:2020tct, Oikonomou:2020sij,Odintsov:2020xji,Odintsov:2020zkl,Oikonomou:2015qha,Haro:2015oqa,Odintsov:2020mkz,Oikonomou:2020oil,Odintsov:2020sqy,Odintsov:2020ilr,Anninos:2019nib}.
In Ref. \cite{Baumann:2015xxa}, it was clearly shown that, at leading order in the  breaking of conformal symmetry, a coupling to the squared Weyl tensor in the gravitational effective action can reproduce the most general higher-curvature corrections to the tensor spectrum, basically breaking the consistency relation between $r$ and $n_{\rm T}$ and possibly leading to blue tensors. However this relation is violated in many other non standard models of inflation and even if a deviation from standard inflation will be observed by future experiments, one may ask how we could convince ourselves that it comes from the higher-curvature effects.

In this work we further investigate higher-curvature corrections at leading order in the breaking of de Sitter isometries. We show that, along with the above mentioned violation of the consistency relation, other non trivial signatures can be left in the tensor two-point function and in particular the running of the tensor tilt can be some order of magnitude larger than expected in the standard slow roll hierarchy~\cite{Giare:2019snj,Martin:2013tda,Zarei:2014bta}, possibly affecting the small scale behavior of tensor perturbations~\cite{Giare:2020vhn,Kuroyanagi:2011iw,Chongchitnan:2006pe,Friedman:2006zt,Smith:2006xf}. If a violation of the slow roll consistency relation were to be observed, a combined analysis of the tilt and the running could in principle shed light on its higher-curvature nature. Finally, we also exploit the possibility of constraining higher-curvature corrections to the inflationary parameters with current and future GWs and CMB data. Indeed properly combining large and small scale measurements the bounds on the tensor tilt and its runnings can be also remarkably improved~\cite{Giare:2020vss}.

The paper is organized as follows. In Sec.~\ref{sec.Spectrum} we compute the higher-curvature corrections to the primordial tensor spectrum showing that an observable violation of the slow roll consistency relation implies also a relatively large tensor running and thus a non trivial scale dependence of the tensor two-point function. In Sec.~\ref{sec.Constraints} we derive constraints on higher-curvature corrections first analyzing the small scales data on gravitational waves and then combining them with the most recent release of CMB data. In Sec.~\ref{sec.Conclusion} we present our conclusions.   

\section{Higher-Curvature Corrections} \label{sec.Spectrum}

At leading order in the breaking of conformal symmetry, the action that reproduces the most general high-curvature corrections to the tensor two-point function reads ~\cite{Baumann:2015xxa,Lee:2017okg} \footnote{Note that a further term $\sim h(\phi) W \tilde W/M^2$ can be considered basically violating parity of primordial tensor modes \cite{Baumann:2015xxa,Lee:2017okg,Lue:1998mq,Alexander:2004wk,Contaldi:2008yz,Takahashi:2009wc}. In our work we ignore such coupling.}

\begin{align}
S= S_{\rm EH} + S_{\phi} + \frac{M_p^2}{2}\, \int d^4x\,\sqrt{-g} \, f(\phi)\, \frac{W^2}{M^2}
\label{action}
\end{align}
where $S_{\rm EH}$ and $S_{\phi}$  are the Einstein-Hilbert action and the action for the Inflaton field $\phi$, respectively. $W$ is the Weyl tensor
\begin{align}
\nonumber W_{\mu \nu \rho \sigma} \doteq  & \nonumber R_{\mu \nu \rho \sigma}\\& \nonumber -\frac{1}{2}\left(g_{\mu \rho} R_{\nu \sigma}-g_{\mu \sigma} R_{\nu \rho}-g_{\nu \rho} R_{\mu \sigma}+g_{\nu \sigma} R_{\mu \rho}\right)\\& +\frac{R}{6}\left(g_{\mu \rho} g_{\nu \sigma}-g_{\nu \rho} g_{\mu \sigma}\right).
\end{align}
involved in the Inflaton-Weyl coupling $f(\phi) W^2 / M^2$ with
\begin{align}
\nonumber  W^{2} &\equiv W^{\mu \nu \rho \sigma} W_{\mu \nu \rho \sigma}\\& =R^{\mu \nu \rho \sigma} R_{\mu \nu \rho \sigma}-2\, R^{\mu \nu} R_{\mu \nu}+\frac{1}{3}\, R^{2}
\end{align}
and $M$ is the scale suppressing higher-curvature corrections. 
Starting from Eq. \eqref{action}, the primordial spectra can be computed to obtain \cite{Baumann:2015xxa}
\begin{equation}
    \mathcal P_{\rm S}(k_*)=\frac{1}{8\pi^2} \frac{H^2}{M_p^2} \frac{1}{\epsilon_1}\frac{1}{\,c_{\rm S}}
\end{equation}
\begin{equation}
    \mathcal P_{\rm T}(k_*)=\frac{2}{\pi^2} \frac{H^2}{M_p^2} \frac{1}{c_{\rm T}}
\end{equation}
where the equations above (as well as all the inflationary parameters) are evaluated at the horizon exit $k_*=(aH)^{-1}=0.05\,\rm{Mpc}^{-1}$. The tensor propagating speed\footnote{The effects of a non trivial propagating speed of PGWs are largely discussed in the literature, \emph{e.g.} Refs.~\cite{Giare:2020vss, Raveri:2014eea,Creminelli:2014wna,Giovannini:2015kfa,Cai:2015ipa,Cai:2015yza,Cai:2016ldn,Fumagalli:2016afy, Gao:2019liu,Noumi:2014zqa,Bordin:2017hal,Bonilla:2019mbm}.}
\begin{equation}
c_{\rm T}\simeq 1-4\left(\frac{H^2}{M^2}\right)f(\phi)
\label{cT}
\end{equation}
is related to the scalar sound speed $c_{\rm S}\simeq 1+ (\epsilon_1 /3) (c_{\rm T}-1)\simeq 1$. Since we are considering the Inflaton-Weyl coupling as a perturbative correction to the gravitational action\footnote{Note that in this way theory is safe from ghost instabilities~\cite{Baumann:2015xxa,Weinberg:2008hq}.}, $c_{\rm T}$ cannot deviate much from unity and this puts constraints on the function $f(\phi)$ and consequently on its scale dependence.  In what follows we consider a simple coupling $d f(\phi)/d\phi \sim \pm 1 / \Lambda$ with $\Lambda < M_p$ and we assume negligible the higher-order derivatives: $d^n f(\phi)/d\phi^n \simeq 0$. We postpone the discussion of a generic coupling-function $f(\phi)$ to appendix \ref{App.A}. Note also that we do not specify the sign of the coupling. In fact, albeit the sign could be constrained by requiring tensor to propagate subliminally, as shown in Refs.~\cite{deRham:2019ctd, deRham:2020zyh} (see also Ref.~\cite{Babichev:2007dw})  this is not always a safe assumption as, depending on the model, it can be possible to perform a change of frame so that in the new frame the tensor speed is $c$, but the speed of the other massless particles is greater than $c $ leaving us with a situation where we have actually constrained the speed of normal species to be superluminal, in tension with causality.

The presence of a non trivial tensor speed breaks the inflationary slow roll consistency relation between $r\simeq 16\epsilon_1/c_{\rm T}\simeq 16\epsilon_1$ and the tensor tilt $n_{\rm t}\simeq -r/8 -\epsilon_{\rm T}$ with $\epsilon_{\rm T}\doteq d\log c_{ \rm T}/d\log k$ \cite{Baumann:2015xxa,Giare:2020vss}. From Eq.~\eqref{cT} it follows that~\cite{Baumann:2015xxa}
\begin{equation}
n_{\rm T}=-\frac{r}{8} + \lambda \, r^{1/2} 
\label{nT}
\end{equation}
where we have ignored negligible terms $\propto (c_{\rm S} -1)$ that are further suppressed by a factor $\epsilon_1$ and we have defined the dimensionless parameter
\begin{equation}
    \lambda \doteq \sqrt{2}M_p \left(\frac{H^2}{M^2}\right) \frac{df(\phi)}{d\phi}\sim \pm \sqrt{2}\left(\frac{M_p}{\Lambda}\right) \left(\frac{H^2}{M^2}\right)
    \label{lambda}
\end{equation}
that weights the size of high-curvature corrections to the inflationary parameters. As discussed in Ref.~\cite{Baumann:2015xxa}, if the inflationary energy scale $H^2$ is close to $M^2$, these corrections can be the dominant effect as the parameter $\lambda$ is also amplified by the factor $M_p/\Lambda$ that can be large. Note also that for enough large positive $\lambda$, higher-curvature corrections can end-up in a blue tensor spectrum, amplifying the PGWs production on the small scales probed by gravitational detectors, as we discuss in Sec.~\ref{sec.Constraints}.

\begin{figure*}[ht!]
    \centering
    \includegraphics[width=.8\textwidth]{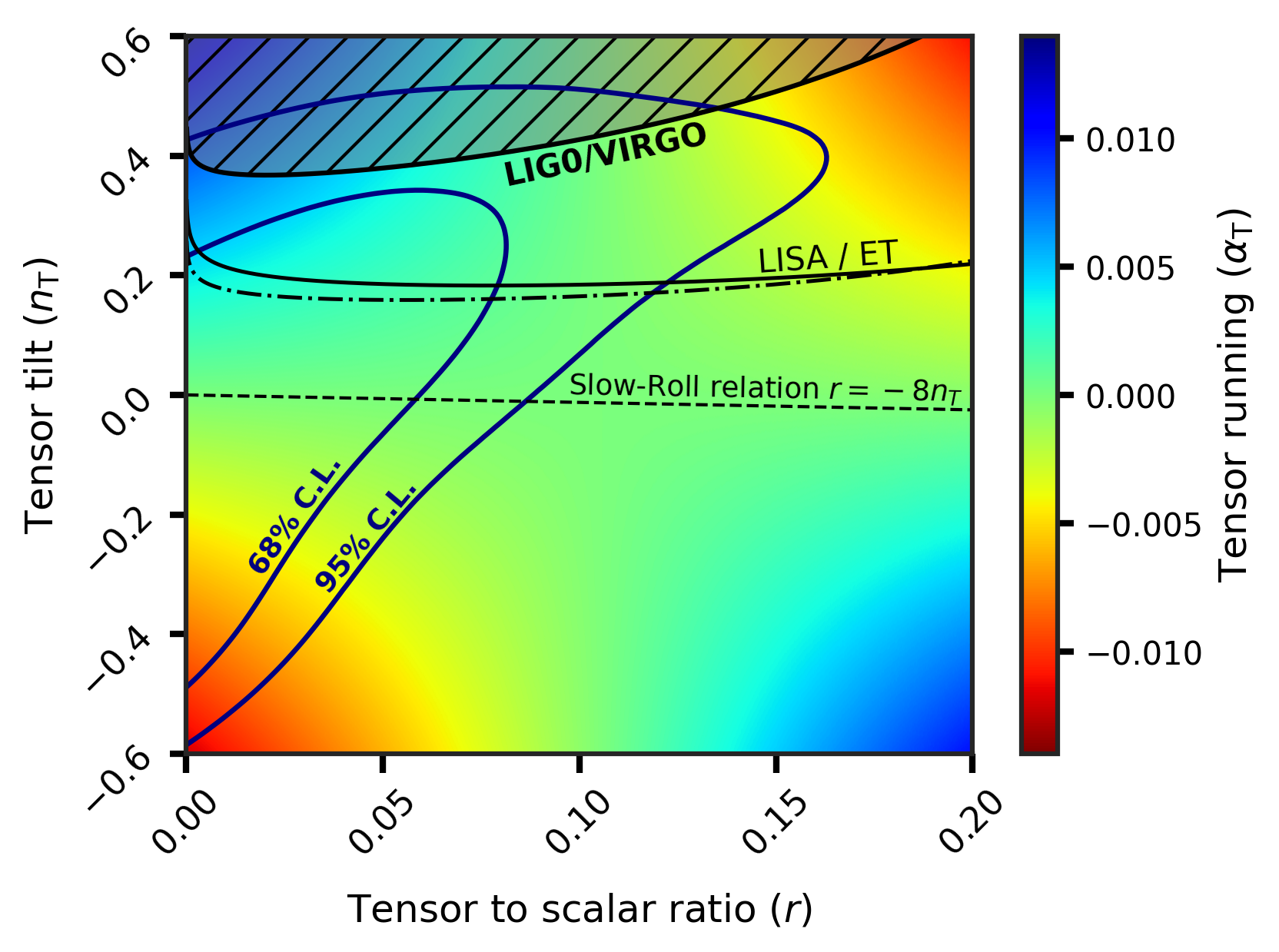}
    \caption{Tensor spectrum expected by higher-curvature corrections. For each point in the plane $\left(r\,,\,n_{\rm T}\right)$ the tensor running $\alpha_{\rm T}$ is fixed by the equations \eqref{nT} and \eqref{alphaT}. The dashed region is excluded by the LIGO/VIRGO limit on the stochastic background \eqref{LVlimit}; the black solid (dashed) line represents the sensitivity expected by LISA (Einstein Telescope). The blue contours are the 68\% and 95\% C.L. bounds for a combination of Planck 2018 \cite{Aghanim:2019ame,Aghanim:2018oex}, BICEP2/Keck 2015~\cite{Ade:2018gkx} and the LIGO/VIRGO~\cite{LIGO_SGWB-2017,LIGO_SGWB-2019} (P18+BK15+LV) data.}
    \label{fig:figure1}
\end{figure*}

We show that along with the tensor tilt, also the other inflationary parameters can acquire non negligible corrections from higher-curvature terms. In particular, by noting that\footnote{We recall the definition of the slow roll parameters $\epsilon_{i\geq 2}\doteq d\log\epsilon_{i-1}/d\log k$, the expression of the scalar tilt $n_{\rm S}-1=-2\epsilon_1-\epsilon_2\simeq -0.04$, its running $\alpha_{\rm S}\doteq d n_{\rm S}/d\log k=-2\epsilon_1\epsilon_2-\epsilon_2\epsilon_3$ and the useful relation $d/d\log k =\sqrt{2}\,M_p\,\epsilon_1^{1/2}\,d/d\phi$.}
\begin{equation}
\frac{d\lambda}{d\log k} = -2\lambda\,\epsilon_1 = -\frac{r}{8}\,\lambda
\label{dlambda/dlogk}
\end{equation}
we derive the expression of the tensor running $\alpha_{\rm T}\doteq dn_{\rm T}/d\log k$, namely
\begin{align}
    \alpha_{\rm T}=\alpha_{\rm T}^{\rm SR}+\lambda\left[-\frac{3}{16}\,r^{3/2} -\frac{1}{2}\,r^{1/2}(n_{\rm S}-1) \right].
    \label{alphaT}
\end{align}
The terms in the square brackets represent the correction with respect to the standard slow roll relation $\alpha_{\rm T}^{\rm SR}=r^2/64+(r/8)(n_{\rm S}-1)$. While in the standard slow roll scenario this relation is $\mathcal O(\epsilon^2)$, implying an extremely small running $\alpha_{\rm T}^{\rm SR}\simeq - 5\times 10^{-n-3}$ for $r\simeq 10^{- n}$, higher-curvature corrections may instead give a relatively large running $\alpha_{\rm T}/\lambda\simeq 2\times 10^{- n/2 - 2}$, see also Fig.~\ref{fig:figure1}.
A large tensor running can leave non trivial features in the shape of the tensor two-point function, affecting the small scale behavior of tensor anisotropies and, if higher-curvature corrections are translated into blue tensors, further enhancing the gravitational wave production on small scales as those probed by gravitational detectors. Therefore if a violation of the consistency relation $r=-8\,n_{\rm T}$ is observed by future CMB and/or small scales measurements, a combined analysis of the tilt and the running should in principle shed light on its higher-curvature nature, see also Fig. \ref{fig:figure1} and the discussion in Sec.~\ref{sec.Constraints.GW}.

As concerns the other inflationary parameters, a computation for the running of running $\beta_{\rm T}\doteq d\alpha_{\rm T}/d\log k$ gives:

\begin{align}
\nonumber \beta_{\rm T}=\beta_{\rm T}^{\rm SR} + \lambda\bigg[&  \frac{15}{256}r^{5/2}+\frac{3}{8}r^{3/2}\,(n_{\rm S}-1) \\& + \frac{1}{4}r^{1/2}\,(n_{\rm S}-1)^2-\frac{1}{2}\,r^{1/2}\,\alpha_{\rm S}\bigg]
\label{betaT}
\end{align}
where $\beta_{\rm T}^{\rm SR}\sim \mathcal O(\epsilon^3)\lesssim 10^{-6}$ represents the standard slow roll term~\cite{Giare:2019snj}. We see that higher-curvature corrections still provide a dominant effect $\beta_{\rm T}/\lambda\simeq 10^{-n/2 - 4}$, which is however extremely small. 

By taking higher order derivatives it is also easy to see that $\alpha^{\rm T}_{j}\doteq (d/d\log k)^j\, n_{\rm T} \lesssim 2^j\,\lambda \times 10^{-\frac{n}{2}-2j}$ from which it follows that the running of order $j+1$ is expected to be a factor $\sim 10^{-2}$ smaller than the running of order $j$.
Despite the fact that higher order runnings can be strongly amplified on ultrahigh $k$, it is easy to see that in this case such terms still remain negligible even on the scales probed by GW interferometers\footnote{We recall that the generic running of order $j$ gives a correction to the tensor tilt that is weighted  by a factor $\log^j(k/k_*)/(j+1)!$ on the generic scale $k$.}. So, along with the tensor tilt, any relevant correction to the spectrum is captured only by the running $\alpha_{\rm T}$ and eventually the running of running $\beta_{\rm T}$.

\section{Constraints}\label{sec.Constraints}
Along with B-modes polarization, primordial tensor fluctuations may have imprinted also the stochastic background of gravitational waves, the analogous of CMB for gravitational waves~\cite{Caprini_2018}. If higher-curvature corrections are translated into blue tensors, the stochastic background $\Omega_{\rm GW}(k)$ can be strongly amplified on the small scales (high $k$) probed by the gravitational detectors and we can use  data by Gravitational Wave experiments to derive constraints on the inflationary parameters~\cite{Akrami:2018odb,Bartolo:2016ami,Cabass:2015jwe,Stewart:2007fu,Wang:2016tbj}. In this section we first derive constraints on higher-curvature corrections using the small scale data on the stochastic background of GWs and then we combine such information with the current CMB data performing a Monte Carlo Markov Chain (MCMC) analysis. We compute the theoretical model using the latest version of the Boltzmann code \texttt{CAMB}~\cite{Lewis:1999bs,Howlett:2012mh} while we use the python sampler \texttt{Cobaya}~\cite{torrado:2020xyz} to extract cosmological constraints. The posteriors of our parameter space have been explored using the MCMC sampler developed for \texttt{CosmoMC}~\cite{Lewis:2002ah,Lewis:2013hha} and tailored for parameter spaces with a speed hierarchy which also implements the "fast dragging" procedure described in \cite{Neal:2005}. The convergence of the chains obtained with this procedure is tested using the Gelman-Rubin criterium~\cite{Gelman:1992zz} and we choose as a threshold for chain convergence $R-1 \lesssim 0.01 $.  To compare current data with the theoretical model, we employ the Planck's 2018 temperature and polarization likelihood which also includes low multipoles data ($\ell < 30$)~\cite{Aghanim:2019ame} combined with the lensing likelihood of Planck's 2018 data release based on temperature and polarization lensing reconstruction~\cite{Aghanim:2018oex} and the CMB power spectrum likelihood of BICEP2/Keck Array (BK15)~\cite{Ade:2018gkx}.

\begin{table*}[ht!]
\begin{center}
\renewcommand{\arraystretch}{1.4}
\begin{tabular}{c@{\hspace{1 cm}}@{\hspace{1 cm}} c @{\hspace{1 cm}} c}
\hline
\textbf{Parameter}                    & \textbf{Prior/Derived} & \textbf{Constraints (P18+BK15+LV)}\\
\hline\hline
$\Omega_{\rm b} h^2$         & $[0.005\,,\,0.1]$ &$0.02240\pm 0.00015$\\
$\Omega_{\rm c} h^2$       & $[0.001\,,\,0.99]$ & $0.1200\pm 0.0012$\\
$100\,\theta_{\rm {MC}}$             & $[0.5\,,\,10]$ & $1.04091\pm 0.00031$\\
$\tau$                       & $[0.01\,,\,0.8]$ & $0.0564\pm 0.0078$ \\
$\log(10^{10}A_{\rm S})$         & $[1.61\,,\,3.91]$ & $3.050\pm 0.015$ \\
$n_S$                        & $[0.8\,,\, 1.2]$ &$0.9653\pm 0.0044$\\
$\epsilon_3$ & $[-0.5\,,\,1]$ & $0.12\pm 0.23 $\\
$r$ & $[0 \,,\,1]$ & $< 0.123$\\ 
$\epsilon_{\rm T}$ & $[-0.5\,,\,0.5]$ & -\\
$\alpha_S$                        & Derived &$-0.0041^{+0.0077}_{-0.0059}$\\
$n_{\rm T}$ & Derived &$0.08^{+0.28}_{-0.19}$ \\
$\alpha_{\rm T}$ & Derived &  $-0.0004^{+0.0031}_{-0.0020}$\\
$\beta_{\rm T}$ & Derived &  $-0.00022^{+0.00084}_{-0.00042}$\\
$\lambda$ & Derived & $0.1^{+2.0}_{-1.2}$\\

\bottomrule
\end{tabular}
\end{center}
\caption{The external priors used in our MCMC sampling and the results obtained combining the full Planck 2018 likelihood~\cite{Aghanim:2019ame,Aghanim:2018oex}, the BICEP2/Keck 2015 B-mode~\cite{Ade:2018gkx} likelihood and the LIGO/VIRGO data on the stochastic background~\cite{TheLIGOScientific:2016dpb}. The constraints on parameters are at 1$\sigma$ level ($68\%$ C.L.) while upper bounds are at 2$\sigma$ ($95\%$ C.L.). We indicate as Derived those parameters obtained by the others using the consistency relations.}
\label{Results}
\end{table*}

\subsection{Constraints from Gravitational Waves} \label{sec.Constraints.GW}

The present day fraction of the energy density of the Universe due to primordial tensor modes at a given scale $k=2\pi\,f$ is~\cite{Akrami:2018odb,Bartolo:2016ami,Cabass:2015jwe,Stewart:2007fu}
\begin{equation}\label{omegaGW}
\Omega_{\mathrm{GW}}(k) \doteq \frac{1}{\rho_{c}} \frac{\mathrm{d} \rho_{\mathrm{GW}}}{\mathrm{d} \log k}=\frac{\mathcal P_{\mathrm{T}} (k)} {24 z_{\mathrm{eq}}}
\end{equation}
where $z_{\rm{eq}} \sim 3400 $ is the redshift at the matter-radiation equivalence \cite{Akrami:2018odb} and $\mathcal P_{\rm T }(k)$ is the primordial tensor spectrum at the scale $k$  
\begin{equation}
\mathcal P_{\rm T}(k) \simeq r\,\mathcal P_{\rm S}(k_*) \left(\frac{k}{k_*}\right)^{n_{\rm T} + \frac{\alpha_{\rm T}}{2}\,\log(k/k_*)+\dots}
\label{PT}
\end{equation}
with the scalar amplitude $\mathcal P_{\rm S}(k_*)\simeq 2.1\times 10^{-9}$ and the pivot scale $k_*=0.05\,\rm{Mpc}^{-1}$.
While a direct detection of the stochastic background has not yet been provided~\footnote{Recently, the North American Nanohertz Observatory for Gravitational Waves (NANOGrav) found strong evidences for a stochastic common-spectrum process \cite{Arzoumanian:2020vkk}. Even if this will be confirmed as a first genuine detection of a stochastic background of GWs, its inflationary interpretation will be in tension with BBN bounds~\cite{Vagnozzi:2020gtf} unless we assume a very low reheating temperature~\cite{Bhattacharya:2020lhc,Kuroyanagi:2020sfw}.}, the first and second observing runs of the LIGO/VIRGO collaboration placed an upper bound on its amplitude for the scales $k_{\rm LV} \in \left(1.3\,\rm{-}\,5.5\right)\times 10^{16} \,\rm{Mpc}^{-1}$, namely
\begin{equation}
\Omega_{\rm{GW}} (k_{\rm LV}) \leq 1.7 \times 10^{-7}.
\label{LVlimit}
\end{equation}
at 95\% C.L. \cite{TheLIGOScientific:2016dpb,LIGO_SGWB-2017,LIGO_SGWB-2019}. Imposing the relations \eqref{nT} and \eqref{alphaT} derived in the previous section, we use the LIGO/VIRGO limit \eqref{LVlimit} to derive constraints on higher-curvature corrections. In Fig.~\ref{fig:figure1} we plot the constraints in the plane $\left(r\,,\,n_{\rm T}\right)$ showing that values $n_{\rm T}\gtrsim 0.4$ are excluded by the LIGO/VIRGO limit \eqref{LVlimit}. Note also that these constraints can be easily translated into constraints on the dimensionless parameter $\lambda$, \emph{i.e.} on the size of the higher-curvature corrections. As we said in Sec.~\ref{sec.Spectrum}, a large positive tensor tilt implies a large positive running $\alpha_{\rm T}$ that is completely fixed by the values of $n_{\rm T}$ and $r$ by equations \eqref{nT} and \eqref{alphaT}. If a violation of the slow roll consistency relation is observed, a test of \eqref{alphaT} could in principle shed light on its higher-curvature nature. Testing this relation with current and future CMB measurements could be extremely challenging as the tensor running, even enhanced by some order of magnitude by higher-curvature corrections, clearly gives higher-order corrections to the tensor spectrum on the CMB scales. Nevertheless, if higher-curvature corrections are translated into a sufficiently large blue tilted spectrum, leading to an $\Omega_{\rm GW}$ visible by future GW experiments, combining the CMB and GW data we might strongly improve our constraining power as also discuss in section~\ref{sec.Constraints.CMB}. Indeed always in Fig.~\ref{fig:figure1} we show the sensitivity curves of future gravitational wave experiments~\footnote{We assumed LISA to have a sensitivity to the stochastic background $\Omega_{\rm{GW}}(k_{\rm Lisa})\simeq 1\times10^{-12}$ on scales $k_{\rm Lisa}\approx 1\times10^{13}\,\rm{Mpc^{-1}}$ \cite{Audley:2017drz} while for the Einstein Telescope we assumed a sensitivity of $\Omega_{\rm{GW}}(k_{\rm ET})\simeq 3\times10^{-13}$ on scales $k_{\rm ET}\approx 5\times 10^{15}\,\rm{Mpc^{-1}}$ \cite{Punturo:2010zz}.} such as LISA~\cite{Audley:2017drz} and Einstein Telescope~\cite{Punturo:2010zz}. They are expected to bring the LIGO/VIRGO upper limits down by a factor $\sim 2$ leading to either a detection or to tighter constraints. Because of \eqref{nT} and \eqref{alphaT}, a detection of $\Omega_{\rm GW}$ at a given scale $k$ will immediately fix the parameter $\lambda$ to
\begin{align}
\lambda=\frac{\frac{\ln\left(\frac{24\,z_{\rm eq}\,\Omega_{\rm GW}(k)}{r\,\mathcal P_{\rm S}(k_*)}\right)}{\ln\left(k/k_*\right)} + \frac{r}{8} - \frac{\alpha_{\rm T}^{\rm SR}}{2}\ln(k/k_*)} {r^{1/2} - \left[\frac{3}{16}\,r^{3/2} +\frac{1}{2}\,r^{1/2}(n_{\rm S}-1) \right] \ln(k/k_*) }.
\end{align}
Supposing that future CMB experiments lead to a first detection of the tensor amplitude $r$, we can use measurements of $\Omega_{\rm GW}(k)$ at different scales (\textit{e.g.} the scales probed by LISA and ET) as consistency check for $\lambda$ and so as a test of equations \eqref{nT} and \eqref{alphaT}. 

We conclude this subsection with a final remark: it is well known that the multi-messenger event GW170817~\cite{GBM:2017lvd,Monitor:2017mdv} sets strong bounds on modified gravity theories, constraining $c_{\rm T}-1\lesssim 10^{-15}$. Therefore one could consider the possibility of using this bound to derive constraints on this model. While it is easy to see that adopting the GW170817 limit higher-curvature corrections will be severely suppressed\footnote{Assuming a coupling function $f(\phi)\sim \phi / \Lambda$  with $\phi\lesssim 10^{15}\rm{GeV}$ the GW170817 limit would imply $|\lambda|\simeq 10^{-11}\ll 1$.}, it is also worth noting that the event GW170817 only constrains the propagating speed of gravity in a precise range of frequencies that is far away from the CMB scales. Because of the running in frequency $\epsilon_{\rm T}=d\log c_{\rm T}/d\log k$, we may not simply use the GW170817 bound as it refers different scales, but we can use constraints on $\lambda$ to relate values of $c_{\rm T}$ at different frequencies. 

\subsection{Constraints from Cosmic Microwave Background and Gravitational Waves}\label{sec.Constraints.CMB}

\begin{figure*}[ht!]
    \centering
    \includegraphics[width=1 \textwidth]{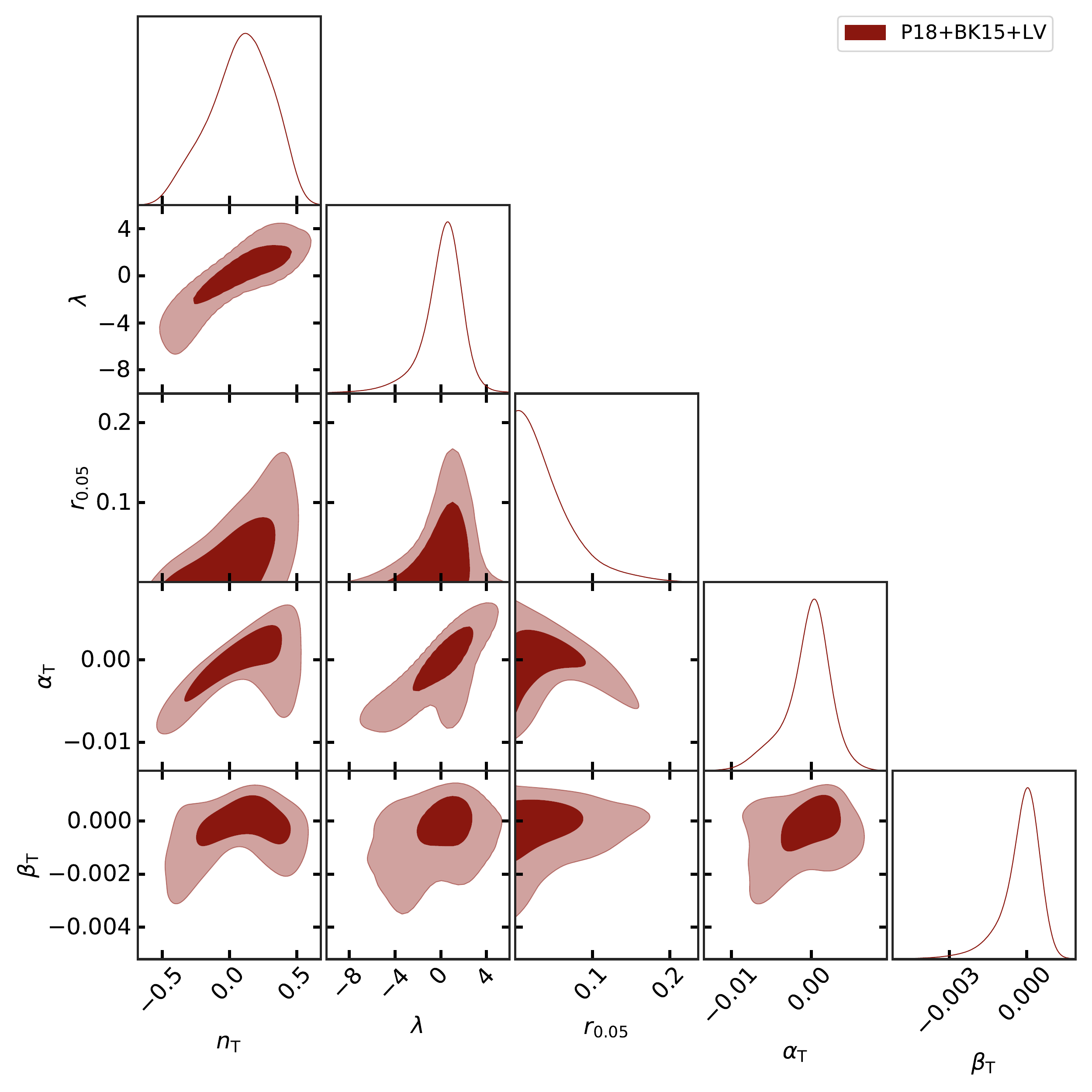}
    \caption{Marginalized 2D and 1D posteriors for the combination of Planck 2018~\cite{Aghanim:2019ame,Aghanim:2018oex}, BICEP2/Keck 2015~\cite{Ade:2018gkx} and the LIGO/VIRGO upper limit on amplitude of the stochastic background~\cite{LIGO_SGWB-2017,LIGO_SGWB-2019} (P18+BK15+LV).}
    \label{fig:figure2}
\end{figure*}

For our MCMC analysis, we consider the six parameters of the standard $\Lambda$CDM model, \textit{i.e.} the baryon $\omega_{\rm b}\doteq\Omega_{\rm b}\,h^2$  and cold dark matter  $\omega_{\rm c}\doteq\Omega_{\rm c}\,h^2$ energy densities, the angular size of the horizon at the last scattering surface $\theta_{\rm MC}$, the optical depth $\tau$, the amplitude of primordial scalar perturbation $\log(10^{10}\,A_{\rm S})$ and the scalar spectral index $n_{\rm S}$. Along with the six standard $\Lambda\rm CDM$ parameters, we also considered the scalar running $\alpha_{\rm S}$, the tensor-to-scalar ratio $r$, the tensor spectral index $n_{\rm T}$, the tensor running $\alpha_{\rm T}$, and the running of running  $\beta_{\rm T}$. However, instead of directly sampling all these parameters (as it is commonly done, see \emph{e.g.}~\cite{Akrami:2018odb,Aghanim:2018eyx}), along with the standard $\Lambda$CDM parameters, we sample only $\{r\,,\epsilon_{3}\,,\epsilon_{\rm T}\}$ and we use the relations derived in Sec.~\ref{sec.Spectrum} to compute the others. More precisely we derive the tensor tilt $n_{\rm T}$ by Eq. \eqref{nT}, its running $\alpha_{\rm T}$ by Eq. \eqref{alphaT} and its running of running $\beta_{\rm T}$ by Eq. \eqref{betaT} with $\alpha_{\rm S}= \alpha_{\rm T}^{\rm SR} +\left( 1 - n_{\rm S}- r/8\right)\epsilon_3$. In this way we are also able to derive constraints on the dimensionless parameters $\lambda$ defined by Eq.\eqref{lambda}, as we discuss below. 

In Table~\ref{Results} we show both the priors used for the sampled parameters, denoting as "Derived" those obtained by consistency relations, and the constraints from the combination of Planck (P18), BICEP2/Keck (BK15) and LIGO/VIRGO (LV) limit on the stochastic background, Eq.\eqref{LVlimit}. We include the LIGO/VIRGO limit as an half-Gaussian prior on the amplitude of tensor spectrum at the smallest scale probed by those gravitational wave interferometers~\footnote{Using Eq. \eqref{omegaGW} the upper bound on the energy density of gravitational waves can be translated into an upper bound on the amplitude of tensor fluctuations at $k_{\rm LV}$. Assuming $z_{\rm eq} \approx 3400$, $\mathcal{P}_{\rm T}(k = 1.3 \times 10^{16} {\rm Mpc^{-1}} ) \leq 1.4 \times 10^{-2}$.}.  In Fig.~\ref{fig:figure2} we instead report the $68\%$ and $95\%$ contour plots for the tensor parameters. 
 
As also discussed in \cite{Giare:2020vss}, the inclusion of the tensor (and scalar) runnings may significantly enhance the constraints on the parameters describing tensor spectra from current data as a large tensor running may affect the small-scale behavior of tensor anisotropies, amplifying the GW power on the ultrahigh $k$ probed by gravitational detectors and possibly leading to an $\Omega_{\rm GW}$ over the LIGO/VIRGO bound~\eqref{LVlimit}. Although our results do not exclude the possibility that observable departures from the slow roll consistency relation can arise from higher-curvature tensors, see also Fig.~\ref{fig:figure1}, they strongly reduce the parameter space allowed for such deviations. In particular our analysis shows a preference for a small running of the tensor tilt $\alpha_{\rm T}=-0.0004^{+0.0031}_{-0.0020}$ at 68\% C.L., consistent with zero as expected in the standard slow-roll hierarchy. The constraints on the tenors running can be translated into a constraint on the dimensionless parameter $\lambda$ that weighs the higher-curvature corrections to the inflationary parameters, namely $\lambda=0.1^{+2}_{-1.2}$ at 68\% C.L., see also Table~\ref{Results} and the discussion in Sec. \ref{sec.Constraints.GW}. 
Also in this case a remarkable preference for values of $\lambda$ consistent with zero is found, disfavoring large corrections from higher curvature tensors, see also the posterior distribution of $\lambda$ in Fig.~\ref{fig:figure2}. Note also that future experiments on GW such as LISA and ET, once combined with current and future CMB data, can further constrain the parameter space allowed for this model. In Fig.~\ref{fig:figure1},  we can appreciate that the sensitivity curves of future gravitational wave experiments intersect the current CMB constraints, which means that a large range of the parameter space currently allowed can be probed by future measurements, leading to either a detection or to tighter bounds on higher-curvature corrections.

\section{Conclusion}\label{sec.Conclusion}

It is well known that several high-energy theoretical models predict higher-curvature corrections to the gravitational effective action. For sufficiently high-scale Inflation, such corrections may leave characteristic signatures in the spectrum of primordial tensor perturbations and a future detection of Primordial Gravitational Waves could therefore open up a unique observational window to test such scenarios. In this paper we further investigate the effects of a coupling of the Inflaton field to higher-curvature tensors in models with a minimal breaking of conformal symmetry. 
In~\cite{Baumann:2015xxa}, it was clearly shown that, at leading order in the breaking conformal symmetry, a coupling to the squared Weyl tensor in the gravitational effective action can reproduce the most general higher-curvature corrections to the tensor spectrum, possibly leading to a violation of  the tensor consistency relation, $n_{\rm T}\ne r/8$, and, depending on the nature of the coupling, to blue tensors. However it is also true that the tensor consistency relation is violated in many other non standard models of inflation and several different plausible mechanisms may predict blue tensors, see \textit{e.g.} \cite{Stewart:2007fu,Capurri:2020qgz}. Therefore even if future gravitational wave and/or CMB measurements will reveal evidence for a violation of the tensor consistency relation, one may ask how we could convince ourselves that it comes from higher-curvature effects. 
In this work we prove that an observable violation of the tensor consistency relation from higher-curvature tensors implies also a running of the tensor tilt some order of magnitude larger than expected in the standard slow-roll hierarchy. A large tensor running may affect the scale dependence of tensor perturbations and, in the case of blue tensors, may amplify the power of the inflationary background of gravitational waves on the small scales probed by gravitational detectors. Deriving a precise relation among the tensor amplitude, the tensor tilt and its running, we show that if a violation of the consistency relation will be observed by future measurements, a test of this relation could in principle shed light on its higher-curvature nature. We also derive and discuss current and future constraints on higher-curvature corrections exploiting current GW and CMB data. We first show that if higher-curvature corrections end-up into blue tensors, the gravitational waves production can be strongly amplified on the small scales probed by the present and future gravitational detectors and that constraints on the amplitude of the stochastic background of gravitational waves can be translated into constraints on the size of higher-curvature corrections, see also Fig.~\ref{fig:figure1}. We then performed a MCMC analysis to compare current data with our theoretical model. In particular we combine the Planck's 2018 temperature and polarization likelihood (which also includes the low multipoles data $\ell < 30$)~\cite{Aghanim:2019ame}, the lensing likelihood of Planck's 2018 data release based on temperature and polarization lensing reconstruction~\cite{Aghanim:2018oex},  the CMB power spectrum likelihood of BICEP2/Keck Array (BK15)~\cite{Ade:2018gkx} and LIGO/VIRGO data on the stochastic background~\cite{TheLIGOScientific:2016dpb,LIGO_SGWB-2017,LIGO_SGWB-2019} that we include as an half-gaussian prior on the tensor amplitude at the LIGO/VIRGO scales. Although our results, shown in Table~\ref{Results}, do not exclude the possibility that observable departures from the slow roll consistency relation can arise from higher-curvature tensors, they constrain the parameter space allowed for such deviations. In particular we found a remarkable preference for a small running of the tensor tilt: $\alpha_{\rm T}=-0.0004^{+0.0031}_{-0.0020}$ at 68\% C.L., which is consistent with what expected in the standard slow roll hierarchy. This is translated into a relatively tight constraint  $\lambda=0.1^{+2}_{-1}$ at 68\% C.L. on the dimensionless parameter $\lambda$, defined by Eq.~\eqref{lambda}, that weighs the size of higher-curvature corrections to the inflationary parameters (with $\lambda=0$ corresponding to the standard slow roll results). We conclude that large corrections from higher curvature tensors, albeit possible, are disfavored by current CMB and GWs data.

\acknowledgments
W.G. and A.M. are supported by "Theoretical Astroparticle Physics" (TAsP), iniziativa specifica INFN. F.R. acknowledges support from the NWO and the Dutch Ministry of Education, Culture and Science (OCW), and from the D-ITP consortium, a program of the NWO that is funded by the OCW.

We thank Sunny Vagnozzi and Antonio Riotto for useful discussions and suggestions.

In this work we made use of the following \texttt{python} packages that are not mentioned in the text : \texttt{SciPy}~\cite{2020SciPy-NMeth} for numerical sampling of the statistical distributions involved in our data analysis, \texttt{GetDist}~\cite{GetDist} a tool for the analysis of MCMC samples which employs \texttt{Matplotlib}~\cite{Matplotlib} for the realization of the plots in the paper and \texttt{NumPy}~\cite{NumPy} for numerical linear algebra.

\appendix

\section{Inflaton coupling function \boldmath{$f(\phi)$}}\label{App.A}

\begin{figure}[htp!]
	\centering
	\includegraphics[width=0.5 \textwidth]{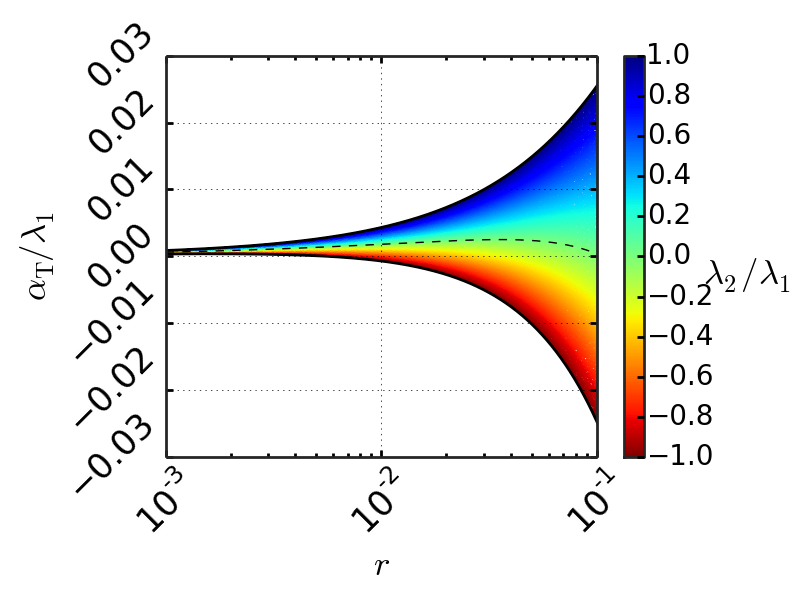}
	\caption{Tensor running for a generic coupling $f(\phi)$. The dashed line represents the model adopted in the paper.}
	\label{fig:figureA.3}
\end{figure}

In this work we have studied higher-curvature corrections to the inflationary parameters considering a coupling between the Weyl tensor and the Inflaton of the form $df(\phi)/d\phi \sim \pm 1/\Lambda$, assuming negligible the higher-order derivatives: $d^n f(\phi)/d\phi^n \simeq 0$. In this appendix we want to generalize our computation for a generic function $f(\phi)$. Introducing the dimensionless parameters
\begin{equation}
    \lambda_n \doteq \left(\sqrt{2}M_p\right)^n \left(\frac{H^2}{M^2}\right) \left(\frac{d}{d\phi}\right)^n f(\phi)
    \label{lambda_n}
\end{equation}
that generalize Eq. \eqref{lambda} with $\lambda_1\equiv\lambda$, we see that Eq. \eqref{dlambda/dlogk} is generalized to
\begin{align}
\frac{d\lambda_n}{d\log k} &= -2\lambda_n\,\epsilon_1 + \lambda_{n+1}\,\epsilon_1^{1/2} \\&= -\frac{1}{8}\,\lambda_n \,r + \frac{1}{4}\,\lambda_{n+1} \, r^{1/2}
\end{align}
So for a generic function $f(\phi)$, while the tensor tilt $n_{\rm T}=-2\epsilon_1-\epsilon_{\rm T}$ is always given by Eq. \eqref{nT}, the tensor running becomes
\begin{align}
    \alpha_{\rm T}=\alpha_{\rm T}^{\rm SR} +\left[-\frac{3\lambda_1}{16}\,r^{3/2} -\frac{\lambda_1}{2}\,r^{1/2}(n_{\rm S}-1) + \frac{\lambda_2}{4}\,r\right].
\end{align}
It differs from Eq. \eqref{alphaT} by a further term $(\lambda_2/4) r$ that can give appreciable contribution only if $|\lambda_2|\simeq |\lambda_1|$. Because of Eq. \eqref{lambda_n}, this means a coupling function of the form $f(\phi)\propto e^{\pm \phi/M_P}$. However in this case we have a further enhancement of the running of tensor tilt, see also Fig.~\ref{fig:figureA.3}. This scenario is even more disfavored by our results that instead show a preference for small running, as we discussed in Sec.~\ref{sec.Constraints.CMB}.
\newpage
\bibliography{main.bib}
\end{document}